\documentclass[aps,twocolumn,prl,amsmath,amssymb,floatfix,superscriptaddress]{revtex4-1}

\usepackage{graphicx}
\usepackage{physics,bm}
\usepackage{color}
\usepackage{lineno}
\begin{document}

\title{Information retrieval and eigenstates coalescence in a non-Hermitian quantum system with anti-$\mathcal{PT}$ symmetry} 

\author{Liangyu Ding}
\author{Kaiye Shi}
\author{Yuxin Wang}
\author{Qiuxin Zhang}
\author{Chenhao Zhu}
\author{Ludan Zhang}
\author{Jiaqi Yi}
\affiliation{Department of Physics, Renmin University of China, Beijing 100872, China}
\author{Shuaining Zhang}
\author{Xiang Zhang}
\thanks{siang.zhang@ruc.edu.cn}
\author{Wei Zhang}
\thanks{wzhangl@ruc.edu.cn}
\affiliation{Department of Physics, Renmin University of China, Beijing 100872, China}
\affiliation{Beijing Academy of Quantum Information Sciences, Beijing 100193, China}
\affiliation{Beijing Key Laboratory of Opto-electronic Functional Materials and Micro-nano Devices, Renmin University of China, Beijing 100872,China}

\begin{abstract}
Non-Hermitian systems with parity-time reversal ($\mathcal{PT}$) or anti-$\mathcal{PT}$ symmetry have attracted a wide range of interest owing to their unique characteristics and counterintuitive phenomena. One of the most extraordinary features is the presence of an exception point (EP), across which a phase transition with spontaneously broken $\mathcal{PT}$ symmetry takes place. We implement a Floquet Hamiltonian of a single qubit with anti-$\mathcal{PT}$ symmetry by periodically driving a dissipative quantum system of a single trapped ion. With stroboscopic emission and quantum state tomography, we obtain the time evolution of density matrix for an arbitrary initial state, and directly demonstrate information retrieval, eigenstates coalescence, and topological energy spectra as unique features of non-Hermitian systems. 
\end{abstract}
\maketitle 


In the framework of quantum mechanics, the Hamiltonian of a closed system is postulated Hermitian, to ensure that the eigenvalues are real and the eigenvectors assemble an orthogonal basis. For an open system interacting with environment, one can obtain an effective non-Hermitian Hamiltonian in some circumstances, where complex eigenvalues and non-orthogonal eigenvectors are commonly expected~\cite{Carmichael1993,Verstraete2009,Diehl2011}. However, in 1998, Bender and Boettcher~\cite{Bender1998} found that a class of non-Hermitian Hamiltonian with parity-time reversal ($\mathcal{PT}$) symmetry can also have real energy spectrum if the $\mathcal{PT}$ symmetry is not spontaneously broken. This work has inspired quite amount of theoretical effort that extends Hermitian quantum mechanics to non-Hermitian systems~\cite{Bender2002,Bender2010}, where exotic physical phenomena were observed near the $\mathcal{PT}$ symmetry breaking point, or usually referred as the exception point (EP)~\cite{El-Ganainy2018,Ozdemir2019,Miri2019}. Owing to the mathematical isomorphism between Schr$\mathrm{\ddot{o}}$dinger equation and classical wave equation, exotic properties of $\mathcal{PT}$-symmetric systems are studied in wave dynamics of optics~\cite{Makris2008,Ruter2010,Chang2014,Brandstetter2014,Feng2014}, active LRC circuits~\cite{Schindler2011}, plasmonics~\cite{Alaeian2014}, acoustics~\cite{Zhu2014}, and microwaves~\cite{Bittner2012}. 

In addition to $\mathcal{PT}$-symmetric Hamiltonian, EP can also present in a system acquiring anti-$\mathcal{PT}$ symmetry~\cite{Peng2016,Choi2018,Ge2013,Yang2017,Wen2020,Wu2014,Jiang2019,Bergman2021}. Since there is a simple relationship between a $\mathcal{PT}$-symmetric Hamiltonian ($\hat{H}^{\rm PT}$) and an anti-$\mathcal{PT}$-symmetric one ($\hat{H}^{\rm APT}$), $\hat{H}^{\rm PT}=\pm i\hat{H}^{\rm APT}$~\cite{Peng2016,Choi2018}, the properties of $\mathcal{PT}$-symmetric systems can find their correspondence in anti-$\mathcal{PT}$-symmetric counterparts. Experimental simulation of anti-$\mathcal{PT}$-symmetric Hamiltonian has also been successfully implemented in flying atoms~\cite{Peng2016}, electrical circuit resonators~\cite{Choi2018}, quantum circuit simulator of nuclear spin~\cite{Wen2020}, and optical systems~\cite{Jiang2019,Bergman2021}.

Among all unique features of $\mathcal{PT}$- or anti-$\mathcal{PT}$-symmetric Hamiltonian, the most exotic and profound ones are the non-unitary reduction of eigenstates~\cite{Kato1995,Gao2015,Ashida2020} and the memory effect in quantum dynamics~\cite{Wen2020, Wang2020,Kawabata2017}. By approaching the EP, the eigenstates associated with different eigenvalues become non-orthogonal and eventually coalesce at the transition point. This observation thus indicates the existence of an entangled partner in the environment if one extends the Hilbert space by including the environment to resemble a Hermitian system. As a result, the information of the system can be stored in the entanglement with environment and flow reversibly upon time evolution, leading to a {\it non-Markovian} dynamics. Apparently, an experimental demonstration of this exotic eigenstates coalescence and information retrieval need a real quantum system and a full characterization of dynamical evolution, which are certainly beyond the scope of simulation of classical systems. As previous experiments have shown the information back flow via quantum simulators~\cite{Wen2020, Wang2020}, a direct witness of such effects in a quantum system with $\mathcal{PT}$ or anti-$\mathcal{PT}$ symmetry is still lacking. 

Here, we demonstrate the first implementation of a quantum system with anti-$\mathcal{PT}$ symmetry and explicitly show the coalescence of eigenstates and information retrieval. Based on a quantum two-level system with effective gain and dissipation~\cite{Naghiloo2019,Li2019,Wu2019,Ding2021,Wang2021}, we impose a periodic driving and realize an effective Floquet Hamiltonian with anti-$\mathcal{PT}$ symmetry~\cite{Bukov2015,Eckardt2017}. Thanks to the quantum nature of our system, we can measure the time evolution of density matrix starting from an arbitrary initial state by means of stroboscopic emission and quantum state tomography~\cite{Leibfried1996}. This ability not only allows us to determine the position of EP, but also provides direct evidence of information flow and eigenstates coalescence. In particular, we find that the von Neumann entropy of the system presents an oscillatory behavior in the anti-$\mathcal{PT}$ symmetry broken region, indicating that the information flows back and forth between the system and the environment~\cite{Kawabata2017}. As a comparison, the entropy monotonically decreases to zero upon time evolution in the anti-$\mathcal{PT}$ symmetry preserving phase. The observation reveals a transition from Markovian to non-Markovian by breaking the anti-$\mathcal{PT}$ symmetry~\cite{Wen2020}. Besides, we find that in the anti-$\mathcal{PT}$ symmetry preserving region, the information of eigenstates can be extracted from the dynamic evolution of density matrix, and the eigenstates coalescence can be directly witnessed by approaching EP. Further, we introduce another tuning parameter of inter-spin detuning, and map out the eigenenergy spectra in parameter space to observe the non-trivial topological structure of the Riemmanian sheets intersecting at EP~\cite{Gao2015}. This non-trivial topology shows that EP is a branch-point singularity~\cite{Ashida2020,Kato1995}.


The experiment is performed on a single trapped $^{171}$Yb$^+$ ion under periodic driving, which is formulated as a dissipative quantum two-level Hamiltonian~\cite{Ding2021}
\begin{equation}
\label{eq1}
\hat{H}(t)=J(t) e^{-i \phi(t)\hat{\sigma}_z}\hat{\sigma}_x-2i\Gamma(t)\dyad{\uparrow}{\uparrow},
\end{equation}
where $\hat{\sigma}_x=\dyad{\uparrow}{\downarrow}+\dyad{\downarrow}{\uparrow}$ and $\hat{\sigma}_z=\dyad{\uparrow}{\uparrow}-\dyad{\downarrow}{\downarrow}$ are Pauli operators in the pseudo-spin space spanned by two hyperfine states $\ket{F=1,m=0}$ ($\ket{\uparrow}$) and $\ket{F=0,m=0}$ ($\ket{\downarrow}$) of the $^{2}S_{1/2}$ manifold, $J(t)$ is the inter-spin coupling strength, and $2\Gamma(t)$ is the decay rate of the $\ket{\uparrow}$ state. The level diagram and experimental setup are illustrated in Figs.~\ref{fig1}(a) and \ref{fig1}(b). Detailed information of the experimental system are elaborated in Supplemental Materials.

\begin{figure}[tbp]
\centering
\includegraphics[width=1\linewidth]{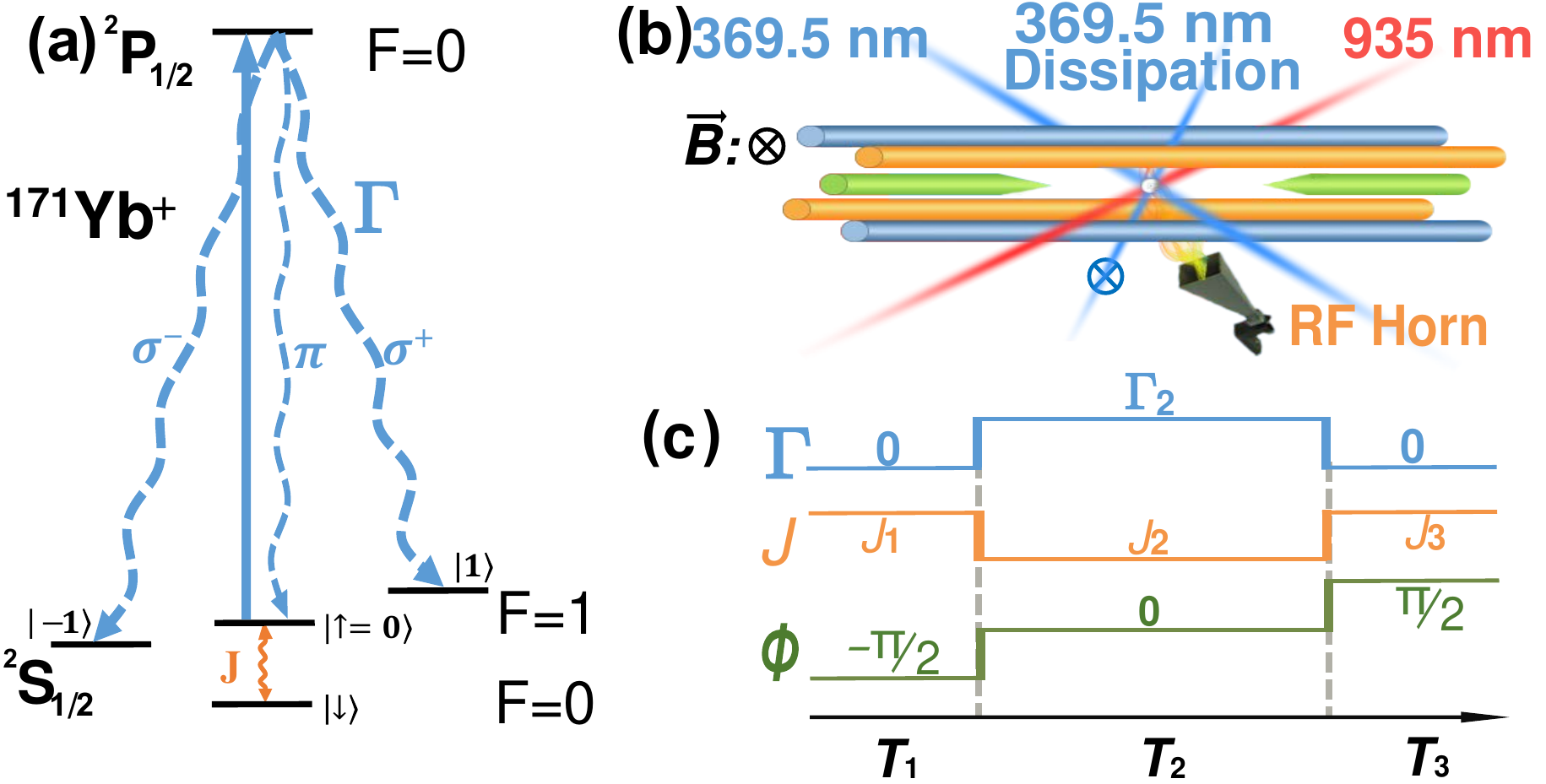}
\caption{Experimental setup. (a) The level diagram of $^{171}$Yb$^{+}$. Blue curves demonstrate the scheme to implement dissipation $\Gamma$ on the state $\ket{\uparrow}$. The state $\ket{\uparrow}$ is stimulated to the upper manifold by a 369.5 nm laser (blue solid line) and then spontaneously decayed back to the ground state manifold with $\pi$ and $\sigma^{\pm}$ polarizations (blue dashed curves). The inter-spin coupling is realized by a microwave of 12.6 GHz (orange curve).
(b) The experimental setup. The ion is trapped in a four-rod trap controlled by lasers and microwaves. The symbol $\otimes$ indicates the direction of magnetic field $\vec{B}$ and the polarization of dissipation beam. RF Horn is used for amplifying and transmitting the coupling microwave. 
(c) The periodic driven scheme for realizing anti-$\mathcal{PT}$ symmetry. $\Gamma$, $J$ and $\phi$ in this figure are in different vertical coordinates.
}
\label{fig1} 
\end{figure}

When the Hamiltonian is subject to a periodic driving with period $T$, i.e., $\hat{H}(t)=\hat{H}(t+T)$, the long-term dynamics is characterized by the evolution operator (with natural unit $\hbar=1$) $\hat{U}_T=\mathcal{\hat{T}}e^{-i\int_0^{T}\dd t \hat{H}(t)}$, where $\mathcal{\hat{T}}$ is the time-ordered operator~\cite{Shirley1965}. The effective Floquet Hamiltonian $\hat{H}_F$ is defined by $\hat{H}_F=i\log{(\hat{U}_T)}/T$. In our scheme, we implement periodic modulation of the parameters $J(t)$, $\Gamma(t)$ and $\phi(t)$ in Eq.~(\ref{eq1}) with a form of square wave. This scheme can be interpreted as a three-step quenching protocol, as shown in Fig.~\ref{fig1}(c).
The evolution operator thus reads $\hat{U}_T=e^{-i\hat{H}_3T_3}e^{-i\hat{H}_2T_2}e^{-i\hat{H}_1T_1}$, where $\hat{H}_{i}$ and $T_{i}$ are the instantaneous Hamiltonian and the duration of each quenching segment, respectively. By tuning the parameters accordingly, we can obtain an effective Floquet Hamiltonian with anti-$\mathcal{PT}$ symmetry
\begin{equation}
\label{eq:FloqH}
\hat{H}^{\rm APT}_F=-\alpha(J\hat{\sigma}_z+i\Gamma\hat{\sigma}_x+i\Gamma \mathcal{\hat{I}}),
\end{equation}
where $\alpha=T_2/T$ and $\mathcal{\hat{I}}=\dyad{\downarrow}{\downarrow}+\dyad{\uparrow}{\uparrow}$ is the identity operator. Here, $J$ and $\Gamma$ are time-independent parameters determined by the driving protocol. A more detailed description of the parameter settings and a theoretical derivation of the Floquet Hamiltonian can be found in Supplemental Materials.

In the essence of Floquet theory, when we observe the system at the end of each complete period, the dynamical evolution of the original $\hat{H}(t)$ is identical to that of a static system described by the Floquet Hamiltonian $\hat{H}^{\rm APT}_F$, and the phase transition of $\hat{H}(t)$ defined by quasi-energy ~\cite{Sambe1973} can be well captured by the eigenvalue of $\hat{H}^{\rm APT}_F$, written as $E_{\pm}=\alpha(-i\Gamma \pm \sqrt{J^2-\Gamma^2})$. When $\Gamma > J$, the system is in the anti-$\mathcal{PT}$ symmetry preserving phase, both the two eigenvalues $E_{\pm}$ are pure imaginary. When $\Gamma < J$, the system is in the anti-$\mathcal{PT}$ symmetry broken regime, the eigenvalues are complex numbers with the same imaginary part but  opposite real part. It is worth noting that $\hat{H}(t) $ itself is not anti-$\mathcal{PT}$ symmetric. The anti-$\mathcal{PT}$ symmetry of the Floquet Hamiltonian $\hat{H}^{\rm APT}_F$ is induced by periodic driving, hence can also be referred as {\it pseudo}-anti-$\mathcal{PT}$ symmetry as in Ref.~\cite{Luo2013}. 

To analyze the phase transition of anti-$\mathcal{PT}$ symmetry broken, we first prepare the system at the initial state $\ket{\psi, t =0}=\ket{\downarrow}$. The time evolution of the quantum state under periodic driving reads $\ket{\psi,t}=\hat{U}(t)\ket{\psi,t=0}=\mathcal{\hat{T}}e^{-i\int_0^{t}dt'\hat{H}(t')}\ket{\downarrow}$, which reproduces the dynamics of the anti-$\mathcal{PT}$ symmetric Hamiltonian $\ket{\psi,t}=e^{-i\hat{H}^{\rm APT}_{F} t}\ket{\downarrow}$ at integer multiples of the period $t = nT$. Therefore, we can use the stroboscopic method to observe the driving system at $t=nT$, and unveil various properties of the anti-$\mathcal{PT}$ symmetric effective Hamiltonian. The population of the $\ket{\downarrow}$ state $n_{\downarrow}(t)=|\bra{\downarrow}\ket{\psi,t}|^2$ is measured by the standard fluorescence counting rate threshold method, while the other state $\ket{\uparrow}$ is measured by adding appropriate $\pi$ flips~\cite{Bruzewicz2019}. 
\begin{figure}[tbp]
\centering
\includegraphics[width=0.95\linewidth]{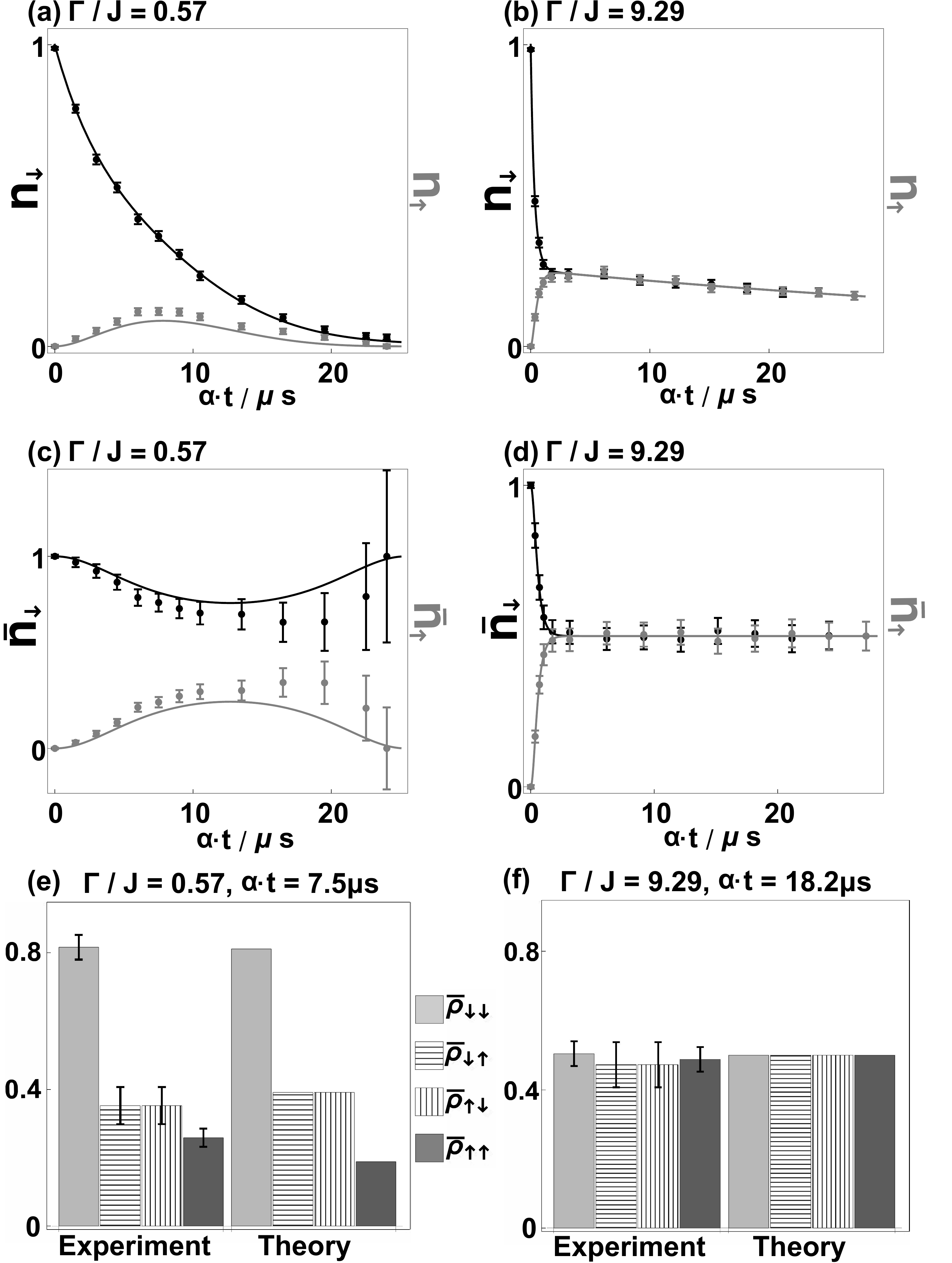}
\caption{(a-b) The evolution of spin state population in the anti-$\mathcal{PT}$ symmetry broken and preserving phases with $\Gamma/J=0.57$ and $\Gamma/J=9.29$, respectively. (c-d) The corresponding normalized populations. (e) The absolute values of  normalized density matrix elements at $\alpha t=7.5$ $\mu$s for the anti-$\mathcal{PT}$ symmetry broken phase. (f) The same as (e) but for the anti-$\mathcal{PT}$ symmetry preserving phase at $\alpha t=18.2$ $\mu$s . For panels (a)-(d), the solid curves are numerical results and the points are experimental data averaged from 1000 rounds of measurement.
}
\label{fig2}
\end{figure} 

The evolution of $n_{\uparrow, \downarrow}(t)$ for two typical sets of parameters are shown in Figs.~\ref{fig2}(a) and \ref{fig2}(b), where the system is prepared in the anti-$\mathcal{PT}$ symmetry broken and anti-$\mathcal{PT}$ symmetry preserving phases, respectively. A stark observation is that the state populations do not present qualitative difference for the two cases. Thus, one can no longer determine the phase transition directly from state populations, as usually done in $\mathcal{PT}$-symmetric cases~\cite{Li2019}. For instead, we analyze the normalized density matrix
\begin{equation}
\label{eq5}
\bar{\hat{\rho}}(t)=\frac{\hat{U}(t)\hat{\rho}(0)\hat{U}(t)^{\dag}}{\mathrm{Tr}[\hat{U}(t)\hat{\rho}(0)\hat{U}(t)^{\dag}]},
\end{equation}
which dominates the dynamics of non-Hermitian Hamiltonian~\cite{Wen2020}. It is a natural generalization of density matrix $\hat{\rho}(t)=\dyad{\psi,t}$ in a Hermitian system~\cite{Kawabata2017}. The diagonal elements of $\bar{\hat{\rho}}(t)$, defined as the normalized population $\bar{n}_{\uparrow, \downarrow}(t)=n_{\uparrow, \downarrow}(t)/[n_{\uparrow}(t)+n_{\downarrow}(t)]$, display discernible difference in different phases. As shown in Figs.~\ref{fig2}(c) and \ref{fig2}(d), $\bar{n}_{\uparrow, \downarrow}(t)$ shows bounded oscillation in the anti-$\mathcal{PT}$ symmetry broken phase, and monotonically tends to $0.5$ in the anti-$\mathcal{PT}$ symmetry preserving phase. Besides, the off-diagonal elements of $\bar{\hat{\rho}}(t)$ can be obtained via state tomography technique, as shown in Figs.~\ref{fig2}(e) and \ref{fig2}(f) for typical evolving times.

\begin{figure}[tbp]
\centering
\includegraphics[width=0.9\linewidth]{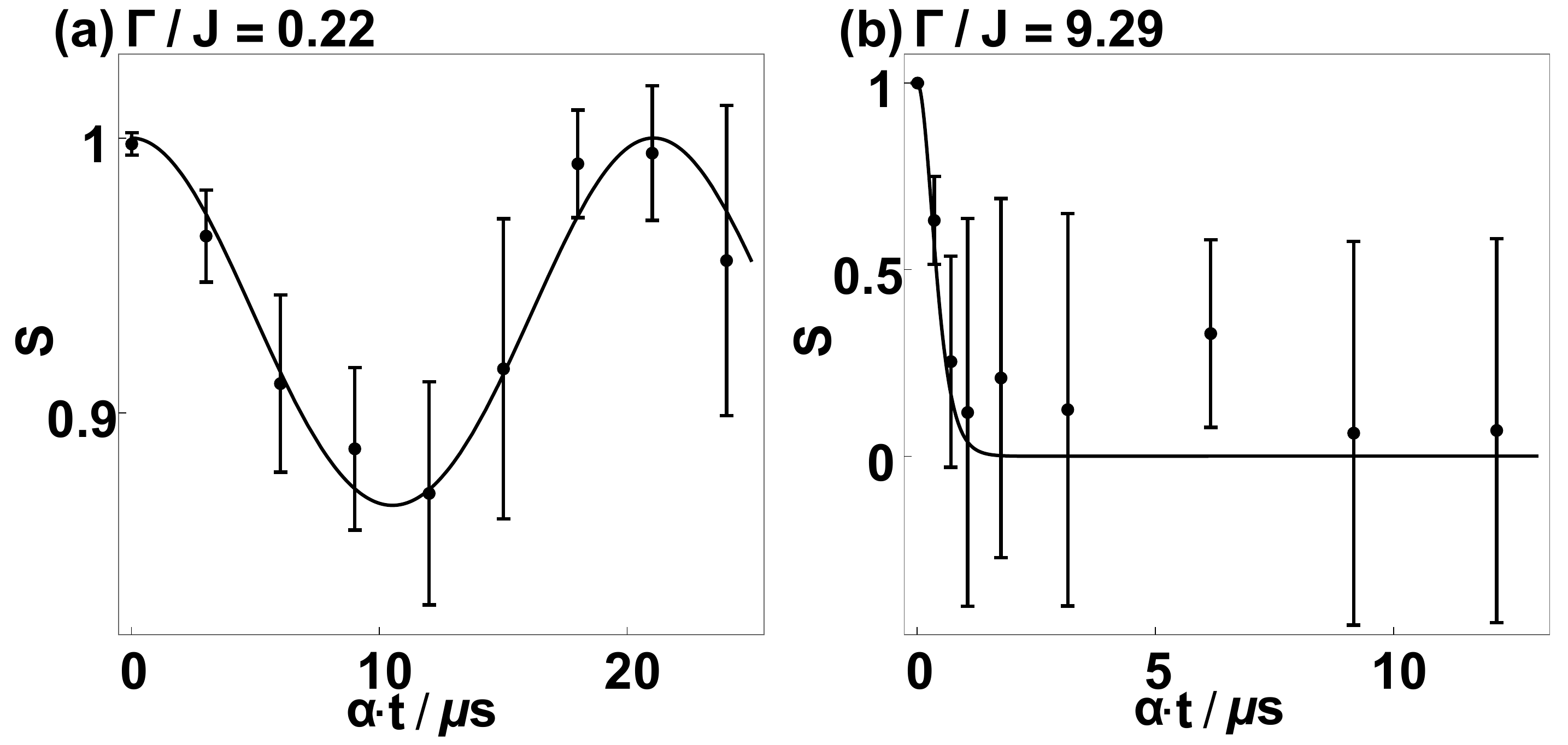}
\caption{(a) The evolution of von Neumann entropy for the anti-$\mathcal{PT}$ symmetry broken phase with $\Gamma/J = 0.22$. The initial state is an equal mixture ($\beta = 0.5$) of $\ket{\psi_1}=(\ket{\uparrow}+\ket{\downarrow})/\sqrt{2}$ and $\ket{\psi_2}=(-\ket{\uparrow}+\ket{\downarrow})/\sqrt{2}$. (b) The same results for the anti-$\mathcal{PT}$ symmetry preserving phase with $\Gamma/J=9.29$. The initial state is an equal mixture of $\ket{\psi_1}=\ket{\downarrow}$ and $\ket{\psi_2}=-\ket{\uparrow}$ states. The solid curves are numerical results and the error bars are calculated by the standard error propagation method.
}
\label{fig3} 
\end{figure}

With the knowledge of density matrix evolution, we can extract the dynamics of von Neumann entropy 
\begin{equation}
S(t)=-\mathrm{Tr}[\bar{\hat{\rho}}(t)\log_2\bar{\hat{\rho}}(t)],
\label{eqv}
\end{equation} 
which describes the flow of information between system and environment. We prepare the qubit at an initial state $\hat{\rho}(0)=\beta\dyad{\psi_1}{\psi_1}+(1-\beta)\dyad{\psi_2}{\psi_2}$, where $\psi_1$ and $\psi_2$ are arbitrary pure states and $\beta$ is a number between 0 and 1. Then the system is evolved and the von Neumann entropy $S(t)$ is calculated by substituting measured results of $\bar{\hat{\rho}}(t)$ into Eq.~(\ref{eqv}). In Fig.~\ref{fig3}, we depict the outcome of $S(t)$ for the two different phases with a proper choice of initial states. Notice that in the anti-$\mathcal{PT}$ symmetry broken region, the entropy presents an oscillatory behavior upon evolution, indicating that the information flows back and forth between system and environment. As a comparison, in the anti-$\mathcal{PT}$ symmetry preserving region, the entropy declines monotonically to zero over time, and the information is lost completely. The information back flow and complete retrieval in the anti-$\mathcal{PT}$ symmetry broken region evidence the unique feature of non-Markovianity of a non-Hermitian system~\cite{Kawabata2017,Breuer2009}, and have potential applications in better controlling of quantum dynamical evolution. 
%


The underlying physics of information back flow is the eigenstates coalescence by approaching EP. In fact, such a phenomenon is in stark contrast to the point of degeneracy in a Hermitian system, also referred as a diabolic point (DP)~\cite{Teller1937}, where the eigenvalues are identical, but the corresponding eigenstates can still be chosen to be orthogonal. Thus, a smoking-gun evidence to distinguish EP from DP is the detection of eigenstates coalescence. With the measurement of density matrix evolution, we can for the first time observe the merging of eigenstates of a quantum non-Hermitian system. 

\begin{figure}[tbp]
\centering
\includegraphics[width=1\linewidth]{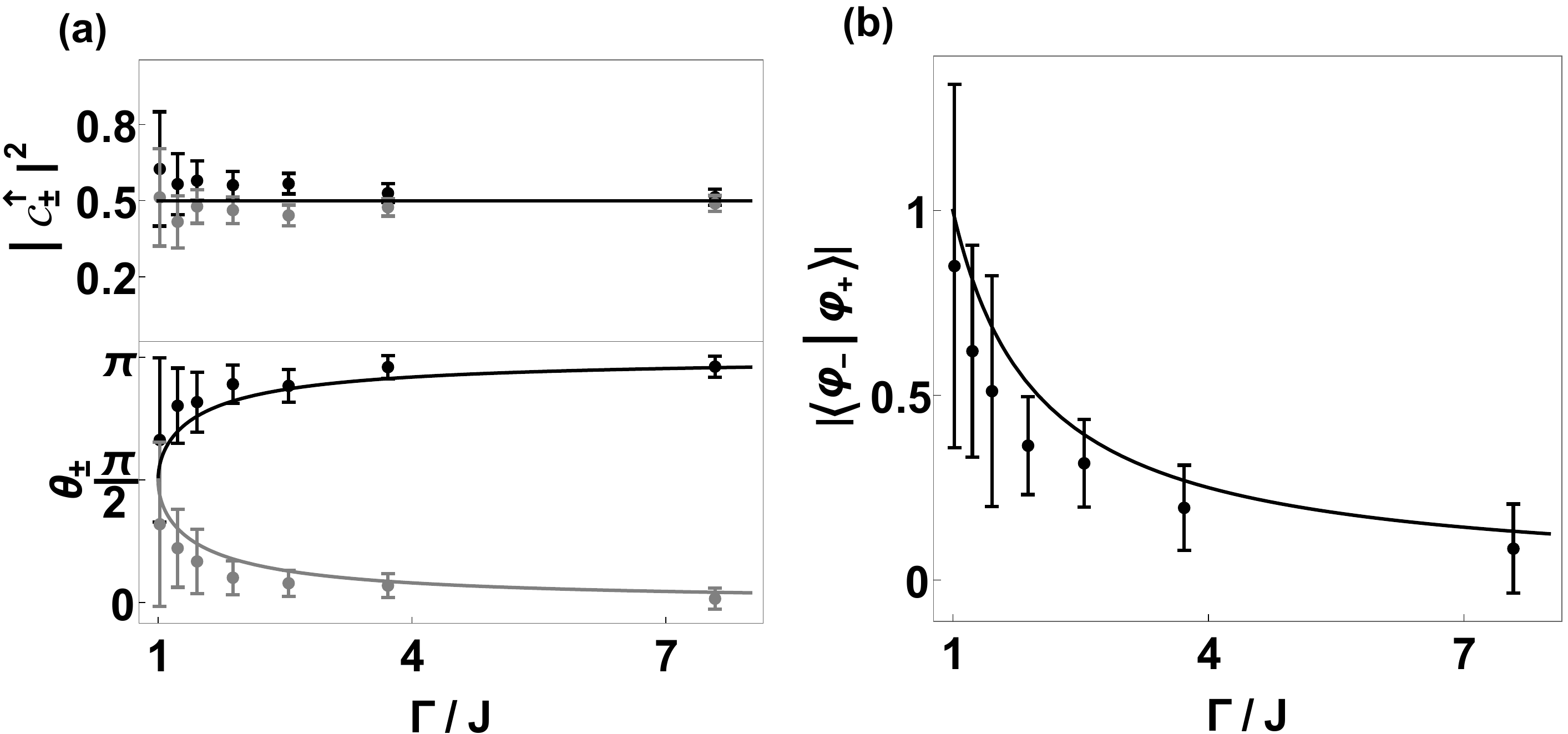}
\caption{(a) The up-spin state population of the eigenstates $|c_\pm^\uparrow|^2$ (top) and the relative phase between the two spin states (bottom). (b) The overlap of the two eigenstates $|\bra{ \varphi_-}\ket{\varphi_+}|$. Numerical simulations are depicted by solid curves.}
\label{fig4} 
\end{figure}

In fact, one can prove that in the anti-$\mathcal{PT}$ symmetry preserving region, the asymptotic limit of the density matrix reads 
$\bar{ \hat{\rho}}(t \to +\infty)=\dyad{\varphi_+}{\varphi_+}$,
where $\ket{\varphi_{+}}$ is one of the two eigenstates of the effective Floquet Hamiltonian Eq.~(\ref{eq:FloqH}). This conclusion suggests that in the anti-$\mathcal{PT}$-symmetric phase, the information of initial state is completely dissipated, such that the steady state is determined solely by the driving Hamiltonian. To extract the other eigenstate $\ket{\varphi_-}$, one can set an inverse phase factor $\phi \to - \phi$ in the $T_1$ and $T_3$ segments within a driving period. This scenario is equivalent to evolving the system reversely, such that a similar expression $\bar{\hat{\rho}}(t \to -\infty)=\dyad{\varphi_-}{\varphi_-}$ can be obtained. The proof of these relations can be found in Supplemental Materials. Although in practice one can not evolve the system for an infinite time, we see from Fig.~\ref{fig2}(f) that all elements of density matrix can approach their limiting values within an experimentally attainable time. 

We set the evolution time as $\alpha t= 15 \mu $s, and scan the effective tunneling parameter $J$ with fixed $\Gamma$ to measure the two density matrices $\dyad{\varphi_\pm}{\varphi_\pm}$, from which the spin projections of the two eigenstates $c_\pm^\uparrow = \bra{ \uparrow}\ket{\varphi_\pm}$ and $ c_\pm^\downarrow = \bra{ \downarrow}\ket{\varphi_\pm}$ can be directly determined. In the upper panel of Fig.~\ref{fig4}(a), we show the population of $|c_\pm^\uparrow|^2$ by varying $\Gamma/J$ in the anti-$\mathcal{PT}$ symmetry preserving phase. Notice that within the entire region of $\Gamma/J >1$, the eigenstates $\ket{\varphi_\pm}$ are equally distributed between the two spin states with $|c_\pm^\uparrow|^2 = |c_\pm^\downarrow|^2 = 0.5$. The experimental data confirms this expectation, although the fluctuation becomes more significant as approaching EP. The lower panel of Fig.~\ref{fig4}(a) depicts the relative phase $\theta_\pm = \textrm{arg}(c_\pm^\downarrow) - \textrm{arg}(c_\pm^\uparrow)$ between the two spin states. With all these information, we can explicitly obtain the overlap of the two eigenstates $|\bra{ \varphi_-}\ket{\varphi_+}|$. As shown in Fig.~\ref{fig4}(b), the overlap tends to unity by approaching EP, ensuring a coalescence of eigenstates of a defective and non-diagonalizable Hamiltonian~\cite{Kato1995}.

\begin{figure}[tbp]
\centering
\includegraphics[width=1\linewidth]{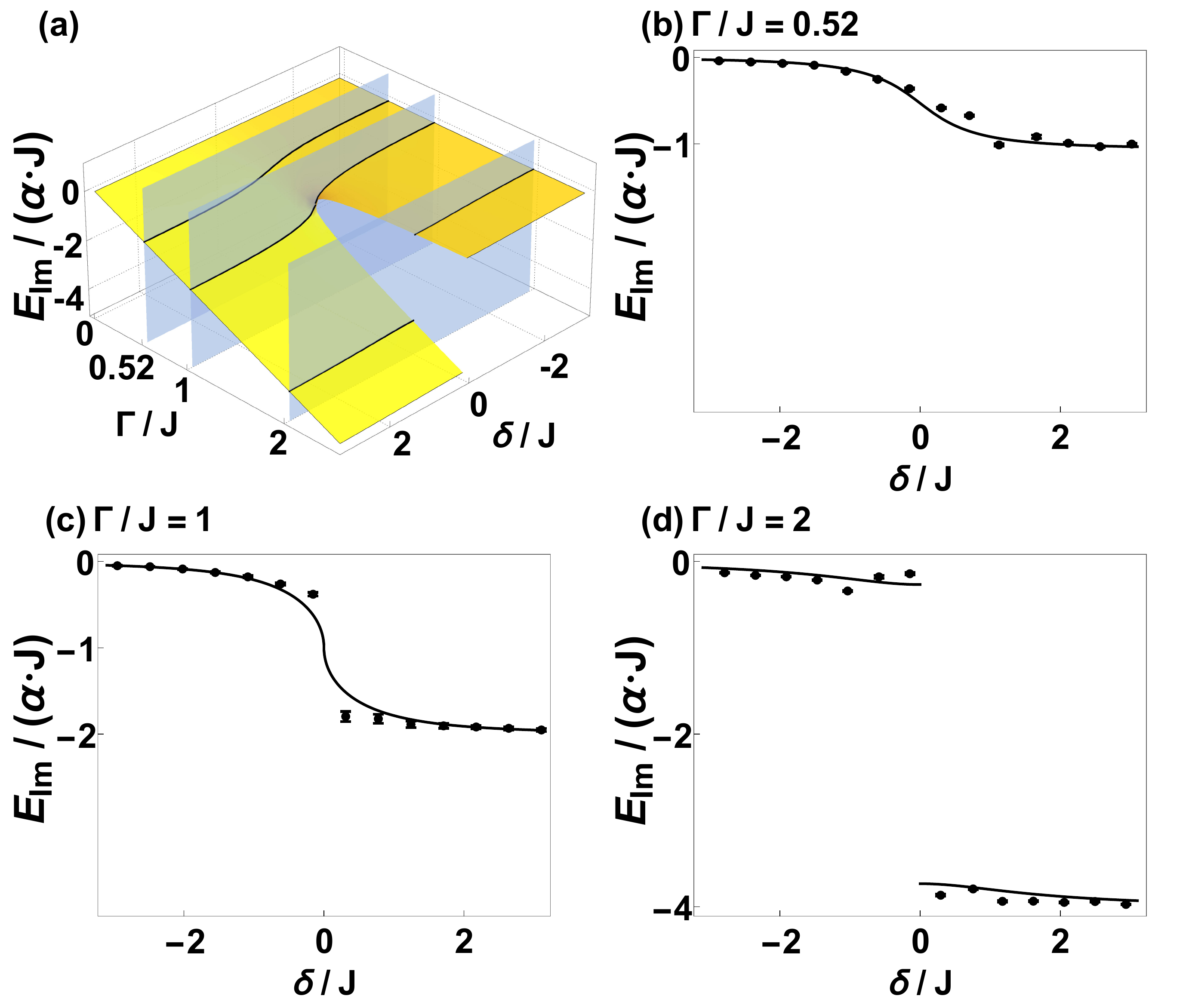}
\caption{Topology of eigenvalues spectra at exception point. (a) The imaginary part of the $E_+$ brach eigenvalue in the parameter space spanned by $\delta/J$ and $\Gamma/J$ (orange surfaces). The three semi-lucent cross sections (gray surfaces) with interaction lines denote a fixed value of $\Gamma/J=0.52$, $1$, and $2$, respectively. (b-d) The numerical results (solid lines) are compared with experimental data (points with error bars) to show the branch-point singularity of EP.}
\label{fig5} 
\end{figure}

The singularity of EP also leads to a non-trivial topological structure of eigenvalue spectra in parameter space. For the two-level Hamiltonian considered here, as the system evolves adiabatically starting from one of its eigenstates, it does not return to the initial state by winding around the EP, but switches to the other eigenstate. To demonstrate this topological characteristic, we introduce an extra $\hat{\sigma}_z$ term in the middle segment $\hat{H}_2$ of the driving sequence via a frequency detuning of the coupling beam. This additional parameter of detuning $\delta/J$, together with $\Gamma/J$, span the parameter space where the Riemann sheets of eigenvalues are defined. The effective Floquet Hamiltonian reads
\begin{equation}
\label{eq7}
\hat{H}^{\delta}_F=\alpha[(\delta -i\Gamma)\hat{\sigma}_x-J\hat{\sigma}_z-i\Gamma \mathcal{\hat{I}}],
\end{equation}
and the eigenvalues are $E^{\delta}_{\pm}=\alpha[ \pm \sqrt{(\delta-i\Gamma)^2+J^2}-i\Gamma]$. In Fig.~\ref{fig5}(a), we show the imaginary part of the $E_+^\delta$ branch by changing $\Gamma/J$ and $\delta/J$ around EP. Apparently, there exists a branch cut starting from EP, such that the system adiabatically evolves from one eigenstate to another by winding around EP~\cite{Bergman2021}. In our experiment, we can extract the eigenvalues by fitting time evolution of the qubit system from the initial state $\ket{\psi, t =0}=\ket{\downarrow}$. By varying $\delta/J$ at three fixed values of $\Gamma/J = 0.52, 1$, and $2$, which correspond to the three sliced sheets in Fig.~\ref{fig5}(a), we show the experimental measurement of $\Im (E^{\delta}_{+})$ in Figs.~\ref{fig5}(b)-(d). A detailed analysis of this protocol and the real part of the $E^{\delta}_{+}$ branch are discussed in Supplemental Materials. 


In summary, we present a realization of anti-$\mathcal{PT}$-symmetric Hamiltonian of a quantum two-level system using a single $^{171}$Yb$^+$ ion. By periodically varying the inter-spin coupling and decay rate, we successfully implement an effective Floquet Hamiltonian with anti-$\mathcal{PT}$ symmetry. The experimental system offers not only an extraordinary controllability of physical parameters, but also the ability to fully describe the dynamical evolution via tomography measurement. Specifically, with the information of density matrix, we can characterize some novel properties of a non-Hermitian quantum system. First, we explicitly demonstrate an information back flow into the system in the anti-$\mathcal{PT}$ symmetry broken phase. We observe an oscillatory behavior of von Neumann entropy, indicating that the system can completely retrieve the information from the environment. This unique non-Markovianity is a direct consequence of the eigenstates coalescence, which is another exotic feature of EP. To show this, we prepare the system in the anti-$\mathcal{PT}$ symmetry preserving phase and obtain the eigenstate wave function from density matrix in the asymptotic limit, which directly reveal the non-orthogonality and coalescence of the eigenstates by approaching EP. We further analyze the system in an extended parameter space, and demonstrate the nontrivial topological structure of eigenvalue spectra around EP. This work shows the great potential of our platform to investigate non-Hermitian quantum systems.

\acknowledgments
L. Ding and K. Shi contributed equally to this work. We thank support from the Beijing Natural Science Foundation (Grant No.~Z180013), the National Natural Science Foundation of China (Grants No.~11522436, No.~11774425, No.~12074428, No.~11704408, and No.~91836106), and the National Key R\&D Program of China (Grant No.~2018YFA0306501).


\end{document}